\documentclass[twocolumn]{jpsj3} %% two-column layout
\usepackage{graphicx}
\usepackage{bm}
\usepackage[usenames]{color}

\title{Novel Interplay between High-$T_c$ Superconductivity and Antiferromagnetism\\ in Tl-based Six-CuO$_2$-Layered Cuprates: $^{205}$Tl- and $^{63}$Cu-NMR Probes}

\author{
Hidekazu Mukuda$^{1}$\thanks{E-mail: mukuda@mp.es.osaka-u.ac.jp}, 
Nozomu Shiki$^{1}$, 
Naoki Kimoto$^{1}$, 
Mitsuharu Yashima$^{1}$, 
Yoshio Kitaoka$^{1}$, \\
Kazuyasu Tokiwa$^{2}$, 
 and 
Akira Iyo$^{3}$
}

\inst{
$^{1}$Graduate School of Engineering Science, Osaka University, Toyonaka, Osaka 560-8531, Japan \\
$^{2}$Department of Applied Electronics, Science University of Tokyo, Noda, Chiba 278-8510, Japan \\
$^{3}$National Institute of Advanced Industrial Science and Technology (AIST), Umezono, Tsukuba 305-8568, Japan}

\abst{
We report $^{63}$Cu- and $^{205}$Tl-NMR studies on six-layered ($n$=6) high-$T_c$ superconducting (SC) cuprate TlBa$_2$Ca$_5$Cu$_6$O$_{14+\delta}$ (Tl1256) with $T_c\sim$100 K, which reveal that antiferromagnetic (AFM) order takes place below $T_{\rm N}\sim$170 K.
In this compound, four underdoped inner CuO$_2$ planes ($n$(IP)=4) sandwiched by two outer planes (OPs) are responsible for the onset of AFM order, whereas the nearly optimally-doped OPs responsible for the onset of bulk SC.
It is pointed out that an increase in the out-of-plane magnetic interaction within an intra-unit-cell causes $T_{\rm N}\sim$~45 K for Tl1245 with $n$(IP)=3 to increase to $\sim$170 K  for Tl1256 with $n$(IP)=4. 
It is remarkable that the marked increase in $T_{\rm N}$ and the AFM moments for the IPs does not bring about any reduction in $T_c$, since $T_c\sim 100$~K is maintained for both compounds with nearly optimally doped OP. 
We highlight the fact that  the SC order for $n\ge5$ is mostly dominated by the long-range in-plane SC correlation even in the multilayered structure, which is insensitive to the magnitude of $T_{\rm N}$ and the AFM moments at the IPs or the AFM interaction among the IPs. 
These results demonstrate a novel interplay between the SC and AFM orders when the charge imbalance between the IPs and OP is significantly large. 
}

\recdate{\today}

\newpage
\begin{document}
\maketitle
%\pacs{74.70.Xa, 74.25.Ha, 76.60.-k}

%\section{Introduction}================================

The parent materials of hole-doped high-$T_{\rm c}$ cuprates are antiferromagnetic (AFM) Mott insulators characterized by a large in-plane superexchange interaction $J_{\rm in}\sim$ 1300 K
%0.1-0.15 eV 
among the $S$=1/2 spins at the Cu sites.
%, since a magnitude $t$ of hybridization in the Cu-O-Cu covalent bonding is large. 
%This is because that a magnitude $t$ of hybridization in the Cu-O-Cu covalent bonding is as large as $t\simeq$1-1.5~eV\cite{review}. 
Since the theory demonstrates that no long-range order takes place at finite temperature for two-dimensional (2D) antiferromagnets \cite{Mermin}, the onset of AFM order is actually mediated by three-dimensional (3D) magnetic interactions such as $J_{\rm out}(n)$ among the CuO$_2$ layers within an intra-unit-cell and $J_{\rm CRL}$ among an inter-unit-cell through the blocking charge reservoir layer (CRL). Note that $J_{\rm out}(n)$ and $J_{\rm CRL}$ are much smaller than $J_{\rm in}$. 
As a matter of fact, as the number of CuO$_2$ planes ($n$) increases in the unit cell, the N\'eel temperature ($T_{\rm N}$) of 325 K for single-layered La$_2$CuO$_4$($n$=1) increases to 410 K for bilayered YBCO($n$=2) and to 537 K for infinite-layered (Ca,Sr)CuO$_2$($n$=$\infty$); nevertheless the size of the AFM moment does not depend on $n$~\cite{Tranquada}.
%is almost kept without depending on $n$.

\begin{figure}[tbp]
\centering
\includegraphics[width=6cm]{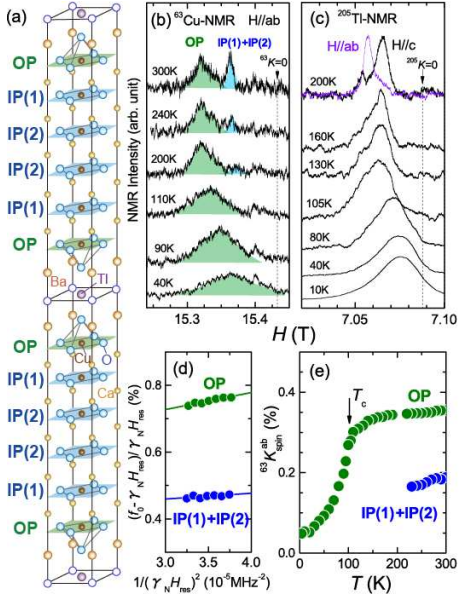}
\caption{(color online) 
(a)  
%Three inequivalent CuO$_2$ layers of six-layered Tl1256 are denoted as OP, IP(1), and IP(2).
Crystal structure of six-layered Tl1256, which includes three inequivalent CuO$_2$ layers denoted as OP, IP(1), and IP(2).  
$T$ dependences of (b) $^{63}$Cu-NMR and (c) $^{205}$Tl-NMR spectra.  (d) Plot of $(f_0-\gamma_{\rm N}H_{\rm res})/\gamma_{\rm N}H_{\rm res}$ against $1/(\gamma_{\rm N}H_{\rm res})^2$ from Eq. (1). 
The slope gives the value of $\nu_{\rm Q}$.  (e) $T$ dependence of $^{63}K_{\rm s}^{ab}(T)$ for OP and IPs. Here, the data at IP(1) and IP(2) are not resolved, presumably due to similar values of the Knight shift and $\nu_{\rm Q}$.
}
\label{f1}
\end{figure}

Multilayered copper oxides TlBa$_2$Ca$_{n-1}$Cu$_{n}$O$_{2n+2+\delta}$ (Tl12($n$-1)$n$) with $n\ge 3$ include two types of CuO$_2$ planes, an outer CuO$_2$ plane (OP) with fivefold oxygen coordination and an inner plane (IP) with fourfold coordination with no apical oxygen, as shown in Fig.~\ref{f1}(a). 
The hole density $p$(OP) is always larger than $p$(IP), since the OP is closer to the CRL than to the IPs. 
In other words, the spatially dependent Madelung potential for hole carriers from the apical oxygen O$^{2-}$ becomes smaller as IPs are apart from the CRL~\cite{Kotegawa2001,Shimizu_all,Kotegawa2004,Mukuda_review}. 
Accordingly, as $n$ increases, the $p$(IP) tends to decrease. 
As a result, the AFM order was observed at underdoped IPs of Tl1245($n$=5) with $T_{\rm N}$=45 K and an AFM moment of  $M_{\rm AFM}\sim$0.1$\mu_{\rm B}$. 
The most important outcome from this result is evidence that the AFM order at each IP  coexists uniformly below $T_{\rm N}$=45 K with the SC state below $T_{\rm c}'$=87 K\cite{Mukuda2008,Mukuda_review}. 
Further reduction of the hole density in the series results in the  uniform coexistence of AFM and SC orders even at an OP where $p$(OP) is less than the critical hole density $p_{\rm c}$ for the onset of AFM order\cite{MukudaPRL2006,Shimizu_all,Mukuda_review}. 
%which removes their misunderstanding that IPs are only responsible for the onset of AFM order, but not for the onset of SC order-parameter amplitude. 
Regarding the $n$ dependence of $p_{\rm c}$,  previous works revealed that as $n$  increases from $n$= 3 to 4 to 5, 
%in Ba$_2$Ca$_{(n-1)}$Cu$_n$O$_{2n}$(F$_y$O$_{1-y}$)$_2$ without apical oxygens, 
$p_{\rm c}$($n$) increases from 0.075 to 0.08 to 0.10~\cite{Shimizu_all,Mukuda_review}, which demonstrates that the interlayer magnetic interaction $J_{\rm out}(n)$ becomes stronger with increasing $n$. 
It is noteworthy that once an AFM order emerges, $M_{\rm AFM}$ is dependent on the doping level $p$ but independent of $n$, whereas $T_{\rm N}$ depends on $n$ or $J_{\rm out}(n)$. 
These results have enabled us to present the ground-state phase diagram of SC and AFM orders as a function of the hole doping level for  the multilayered cuprates with $3 \!\le \!\!n \!\!\le \!5$ and to compare it with the theoretically deduced phase diagrams for 2D doped Mott insulators~\cite{Shimizu_all,Mukuda_review}. 

In this Letter, we report on $^{63}$Cu- and $^{205}$Tl-NMR studies of the six-layered cuprate Tl1256 with $T_c$$\sim$100 K, which provide evidence that an AFM order takes place below $T_{\rm N}$$\sim$170 K. 
In the homologous series of  Tl12($n$-1)$n$ ($n\le6$) in the optimally doped regime, we find that $T_{\rm N}$ is four times larger for $n$=6 than for $n$=5, whereas no trace of AFM order was observed for  $n\le 4$. 
This is considered to be because $J_{\rm out}(n)$ for $n$=6  is significantly larger than that  for $n$=5. 
Since $T_{\rm c}$ for both compounds is almost the same, we remark that even in the multilayered cuprates, the SC order is mostly dominated by the long-range in-plane SC correlation, which is insensitive to the magnitude of $T_{\rm N}$ and AFM moments at the IPs. 
This is the unique interplay between  the high-$T_{\rm c}$ SC and AFM orders observed in the six-layered cuprates.
% in comparison with the previous results for $n\le5$. 
%, demonstrating a novel interplay between the SC and AFM orders in doped Mott insulators. 
%Since the $T_{\rm c}$ for both the compounds is almost the same, we suggest that a possible interlayer Josephson coupling between AFM IPs and OP is insensitive to the magnitudes of $T_{\rm N}$ and AFM moments at IPs or the AFM interaction among IPs, demonstrating an intimate interplay between the SC and AFM orders in doped Mott insulators.

%\section{Experimental} =====================================

A polycrystalline sample of TlBa$_2$Ca$_5$Cu$_6$O$_{14+\delta}$ (Tl1256)  was  synthesized under an appropriate pressure and temperature\cite{Iyo_TcVsn}. 
The lattice parameters were $a$=3.846 \AA\ and $c$=25.380 \AA, determined by powder X ray diffraction.  $T_{\rm c}$ was evaluated to be 100 K from the onset of a sharp diamagnetic signal in the dc susceptibility. 
For the NMR measurements, we used oriented powder samples aligned along the $c$-axis under a high magnetic field of $H_0$=15 T. 
The $^{63}$Cu- and $^{205}$Tl-NMR spectra were obtained at a frequency of $f_0$=174.2 MHz by sweeping an external field perpendicular or parallel to the $c$-axis. The present NMR results for Tl1256 ($n$=6) completely differ from those reported thus far for the compounds with $n\le$ 5, although it is not completely ruled out that a negligible  amount of impurity phases such as Tl1245($n$=5) are contaminated in the sample preparation process.

%fig1------------------------------------------------
%------------------------------------------------

%\section{Results and discussion}=============================

Figure~\ref{f1}(b) indicates the temperature ($T$) dependence of the $^{63}$Cu-NMR spectrum for the central transition ($+$1/2$\Leftrightarrow$-1/2). Since the NQR frequency $\nu_{\rm Q}$ of $^{63}$Cu (nuclear spin $I$=3/2) is sufficiently smaller than the Zeeman field, the central peak for the NMR spectrum (+1/2$\Leftrightarrow$-1/2 transition) is shifted by the second-order perturbation of the nuclear quadrupole interaction. In the case that an external field is applied perpendicular to the $c$-axis, the following relation is valid:
\begin{equation}
\frac{f_0-\gamma_{\rm N}H_{\rm res}}{\gamma_{\rm N}H_{\rm res}}=K^{\rm ab}+\frac{3\nu_{\rm Q}^2}{16(1+K^{\rm ab})}\frac{1}{(\gamma_{\rm N}H_{\rm res})^2},
\label{eq:shift}
\end{equation}
where $f_0$ is the NMR frequency, $\gamma_{\rm N}$ the nuclear gyromagnetic ratio of a Cu nucleus, $H_{\rm res}$ the NMR resonance field, $\nu_{\rm Q}$ the nuclear quadrupole frequency, and $K^{\rm ab}$ the Knight shift for $H\perp c$. 
% and $K^{\rm ab}$ the Knight shift with the field perpendicular to the $c$-axis. 
As shown in Fig. \ref{f1}(d), the slope of the plot of $(f_0-\gamma_{\rm N}H_{\rm res})/\gamma_{\rm N}H_{\rm res}$ against $1/(\gamma_{\rm N}H_{\rm res})^2$ gives the value of $\nu_{\rm Q}$ at the IPs and OP. 
The estimated values of $^{63}\nu_{\rm Q}$=9.7 and 16.6 MHz allow us to assign the two peaks in the spectra of Fig.~\ref{f1}(b) to the IPs and OP, respectively, since the similar values of $^{63}\nu_{\rm Q}$ have been  reported for the IPs and OP previously \cite{Kotegawa2004}.
Note here that, as shown in Fig.~\ref{f1}(a), the four IPs include two inequivalent IPs, IP(1) and IP(2), but it is not possible to resolve each spectrum associated with IP(1) and IP(2) owing to their comparable values of $\nu_Q$.

Figure~\ref{f1}(e) indicates the $T$ dependence of the spin component of the $^{63}$Cu Knight shift $^{63}K_{\rm s}^{ab}$, which is evaluated by subtracting the orbital part of the Knight shift $^{63}K_{\rm orb}^{ab}\sim$0.21($\pm$0.01)\% from $K^{\rm ab}$. 
The marked decrease in $^{63}K_{\rm s}^{ab}$ below $T_{\rm c}$ reveals the decrease in the spin susceptibility owing to the formation of a spin-singlet Cooper pairing. 
The empirical relationship between the hole density $p$ and the value of $^{63}K_{\rm s}^{\rm ab}$ at 300 K~\cite{Shimizu_p-Ks} enables us to evaluate $p$(OP) to be $\sim$0.152 and $\overline{p}$(IPs) to be $\sim$0.076. 
Here, the $\overline{p}$(IPs) represents the average value of $p$[IP(1)]  and $p$[IP(2)]. 
Provided that $p$ at each CuO$_2$ plane follows the spatial dependence of the Madelung potential from the apical oxygen O$^{2-}$[see Fig. 2(d)], the relation $p$(OP)$>\!\!\!p$[IP(1)]$>\!\!\!p$[IP(2)] is expected and $p$[IP(1)]$\sim$0.086 and $p$[IP(2)]$\sim$0.07 are tentatively estimated using $p$(OP)=0.152.
The average value $\overline{p}$(IPs) of $p$[IP(1)] and $p$[IP(2)] almost coincides with $\overline{p}$(IPs)$\sim$0.076 extracted experimentally from $K^{\rm ab}$\!\!. 
As possible evidence for the AFM order at IPs, we note that the Cu-NMR spectra at the IPs disappear below $\sim$200 K owing to the development of AFM correlations upon cooling toward $T_{\rm N}$, but the Cu-NMR spectrum at the OP is observable down to low temperatures, suggesting that the OP does not play a primary role in the onset of AFM order.   

%fig1------------------------------------------------
\begin{figure}[tbp]
\centering
\includegraphics[width=6.5cm]{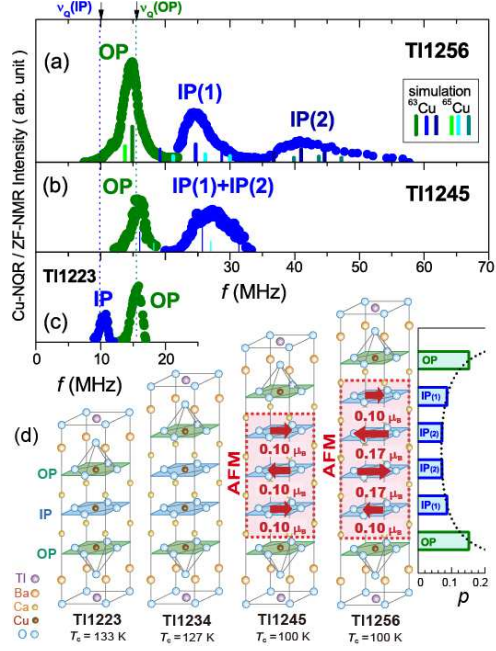}
\caption{(color online)  
Zero-field Cu-NMR/NQR spectra of Tl12($n$-1)$n$ at 1.8 K  for (a) $n$=6, (b) $n$=5 ($T_{\rm c}$=100 K)~\cite{Mukuda_review}, and (c) $n$=3 ($T_c$=133 K). 
The solid bars in (a) and (b) are  simulated  resonance frequencies and intensities for the  two isotopes $^{63}$Cu and $^{65}$Cu (see text). 
Dotted lines represent typical NQR frequencies ($^{63}\nu_{Q}$) in the case that  the IPs and OP are in the nonmagnetic state. (d) Crystal structure of Tl12($n$-1)$n$ with $3\le n\le 6$, schematic views of the AFM order at 1.8 K for $n\ge$5, and the distribution of the hole density at each layer for $n$=6. 
}
\label{f2}
\end{figure} 
%------------------------------------------------

Zero-field Cu-NMR/NQR measurements were performed at 1.8 K for Tl1256($n$=6) to gain an insight into the AFM ordered state.  
Figure~\ref{f2} indicates the NMR spectra for (a) Tl1256, (b) Tl1245($n$=5) with $T_{\rm c}$=100 K,~\cite{Mukuda_review} and (c) Tl1223($n$=3) with $T_c$=133~K. 
The nuclear Hamiltonian ${\cal H_N}$ for the Cu nucleus ($I$=3/2) at zero external field is described in terms of the Zeeman interaction ${\cal H_Z}=-\gamma_{\rm N} \hbar {\bf H_{\rm int}}\cdot{\bf I}$ and the nuclear quadrupole interaction ${\cal H_Q}=(h\nu_Q/2)(I_{z'}^2-5/4)$, where $H_{\rm int}$ is the internal field at the Cu site induced by AFM moments. 
In the case that the IPs and OP are in the nonmagnetic state with $H_{\rm int}$=0, the Cu-NQR spectra should be observed at $^{63}\nu_{\rm Q}$(IPs)$\sim$9.7 and $^{63}\nu_{\rm Q}$(OP)$\sim$16.6 MHz, as shown by the dotted lines in Fig. \ref{f2}. 
As a matter of fact, such the spectra are observed in Fig. 2(c) for Tl1223($n$=3), which  does not have an AFM order. 
Accordingly, the spectrum with a peak at $\sim$16 MHz in Fig. \ref{f2}(a) is assigned to the OP, which is in the paramagnetic state. 
For the IPs, the peaks around $\sim$25 and 42 MHz in the spectra are assigned to IP(1) and IP(2), respectively, revealing that both are in the AFM ordered state.
This is because these frequencies are much higher than $^{63}\nu_{\rm Q}$(IPs)$\sim$9.7 MHz, pointing to the presence of an internal field $H_{\rm int}$ at the IPs. 
In fact, on the basis of the nuclear Hamiltonian of ${\cal H_N}=\cal H_{\rm Z}+\cal H_Q$,
%$=$-\gamma_{\rm N} \hbar {\bf H_{\rm int}}\cdot{\bf I}$ +$(h\nu_Q/2)(I_{z'}^2-5/4)$, 
a simulation of these spectra gives $H_{\rm int}$=2.1 and 3.6 T for IP(1) and IP(2), respectively.
%\textcolor{blue}{
%Note that in the previous studies on $n$=5 compounds\cite{Mukuda_review} the spectra of  IP(1) and IP(2)  were not resolved owing to less discrimination of electronic states at IP(1) and IP(2). 
%}
Using the relation $H_{\rm int}$=$|A_{ab}-4B|M_{\rm AFM}$, spontaneous AFM moments of  $M_{\rm AFM}$(IP(1)) =0.10$\mu_{\rm B}$ and $M_{\rm AFM}$(IP(2))=0.17$\mu_{\rm B}$ are estimated at $T$=1.8 K, as illustrated schematically in Fig. \ref{f2}(d). 
Here, we assume that the on-site hyperfine field is $A_{ab}\approx$ 3.7 T/$\mu_{\rm B}$, the transferred hyperfine field from the nearest-neighbor Cu site is $B({\rm IP})\approx$ 6.1 T/$\mu_{\rm B}$, and $B({\rm OP})\approx$ 7.4 T/$\mu_{\rm B}$~\cite{Kotegawa2004}.   
As a result, the AFM order with $M_{\rm AFM}$(IP(1))=0.10$\mu_{\rm B}$ and $M_{\rm AFM}$(IP(2))=0.17$\mu_{\rm B}$ emerges at the respective hole densities of $p$(IP(1))$\sim$0.086 and $p$(IP(2))$\sim$0.07. 
%\textcolor{blue}{
We highlight the fact that the present relationship between $M_{\rm AFM}$ and $p$ in Tl1256($n$=6) coincides with the ground-state phase diagram of the SC and AFM orders as a function of $p$ for the multilayered cuprates with $3\le n \le 5$\cite{Mukuda_review}, ensuring that this phase diagram is independent of $n$.
%}

%fig3------------------------------------------------
\begin{figure}[tbp]
\centering
\includegraphics[width=6.5cm]{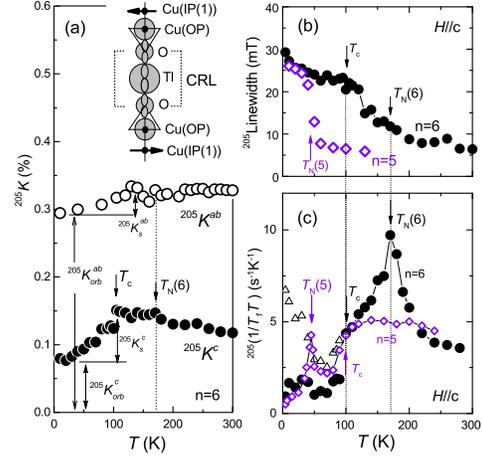}
\caption{(color online) 
 (a) $T$-dependence of $^{205}$Tl Knight shift for $H\|c$ and $H\|ab$. See the text for a description of the inset.
$T$-dependences of (b) $^{205}$Tl-NMR linewidth for $H\|c$ and (c) $^{205}(1/T_1T)$ for Tl1256 and Tl1245. The peaks of $^{205}(1/T_1T)$ indicate that $T_{\rm N}$=170 and 45 K for Tl1256 and Tl1245, respectively. The open diamonds in (b) and (c) represent the $^{205}$Tl-NMR data for Tl1245 with $T_c$=100 K and $T_{\rm N}$=45 K\cite{Mukuda_review}. }
\label{f3}
\end{figure}
%------------------------------------------------

Here, we show the $^{205}$Tl-NMR results at the CRL, which enable us to gain an insight into the 3D magnetic interaction $J_{\rm CRL}$ among the inter-unit-cells through the CRL and determine $T_{\rm N}$. 
Figure \ref{f1}(c) shows the $T$ dependence of the $^{205}$Tl-NMR spectra.  
The spectra for $H\|c$ and $H\|ab$ (see the data at 200 K) are well resolved and the $T$ dependences of $^{205}$Tl Knight shift for $H\|c$ and $H\|ab$ are presented in Fig. \ref{f3}(a).  
$^{205}K^{ab(c)}$ comprises the spin component $^{205}K_{\rm s}^{ab(c)}$ and the $T$-independent orbital component $^{205}K_{\rm orb}^{ab(c)}$~\cite{Kambe}. 
$^{205}K_{\rm s}^{ab(c)}$ is associated with an anisotropic hyperfine field supertransferred by the $6s$- and $6p$-spin polarizations at the Tl site, originating from the spin polarization at the IPs and OP. 
As shown in Fig. \ref{f3}(a), the rapid decrease in $^{205}K^{ab(c)}$ below $T_c$ is attributed to that in $^{205}K_{\rm s}^{ab(c)}$ due to the formation of spin-singlet Cooper pairs. 
Since $^{205}K_{\rm s}^{ab(c)}$ approaches zero well below $T_c$, the values of $^{205}K^{ab}$ and $^{205}K^{c}$ at $T \ll T_c$ give $\sim$0.29 and $\sim$0.07\%, respectively.
Furthermore, it is anticipated that the anisotropic spin part in the Knight shift exhibits $^{205}K_{\rm s}^{c}\!\!>^{205}\!\!K_{\rm s}^{ab}$, suggesting that the dipole hyperfine field from the spin polarization at the $6p_z$ orbital is larger than those at the $6p_{x,y}$ orbitals.
By contrast, the anisotropy in the orbital part exhibits $^{205}K_{\rm orb}^{c}\!\!<^{205}\!\!\!K_{\rm orb}^{ab}$. 
In general, $K_{\rm orb}$ originates from the van Vleck susceptibility, which is proportional to $|\langle g|{\bm L}|e\rangle|^2\!/|\epsilon_g\!-\!\epsilon_e|$. 
Here, $\epsilon_g (\langle g|)$ and $\epsilon_e (|e\rangle)$ denote the energies (wave functions) for the ground state and the excited states, respectively, and ${\bm L}$ is the orbital angular momentum operator\cite{Vyaselev}.
Provided  that $6p_z$ and $6p_{x,y}$ are in the ground and excited states, respectively, it is expected that $^{205}K_{\rm orb}^z \!\!\propto\!\! |\langle p_z|L_z|p_{x,y}\rangle|^2$=0 and $^{205}K_{\rm orb}^{x,y} \!\!\propto\!\! |\langle p_z|L_{x,y} |p_{y,x}\rangle|^2\!\!>$0. 
As a result, the spin densities on the $6s$ and $6p_z$ orbitals at the Tl site are responsible for the anisotropy in the spin and orbital parts, playing a significant role in the interlayer magnetic coupling $J_{\rm CRL}$ through the possible covalency with the $4s$ and $3d_{x^2-y^2}$ orbitals of Cu via the $2p_z$ and $2s$ orbitals of apical O. 
This event is schematically drawn in the inset of Fig. \ref{f3}(a). 
These microscopic outcomes resemble the previous $^{205}$Tl-NMR results for single-layered Tl1201\cite{Kambe,Vyaselev}, indicating that the local electronic state at the Tl site, and hence $J_{\rm CRL}$, does not vary from $n$=1 to 6. 

Next, we estimate $T_{\rm N}$ from the measurements of $^{205}(1/T_1T)$.
Its $T$ dependence exhibits a peak at 170 K, as shown in Fig. \ref{f3}(c).
The distinct peak in $^{205}(1/T_1T)$ gives evidence that an AFM order occurs in a long-range manner.
Below $T_{\rm N}$, as shown in Fig. \ref{f3}(b), the linewidth in the $^{205}$Tl-NMR spectra starts to increase, similar to the case for Tl1245, giving another evidence for the onset of the AFM order.  
These similar behaviors were observed in the $^{205}$Tl-NMR spectra of Tl1245($n$=5) at $T_{\rm N}$(5)=45 K, below which the underdoped IPs with $\overline{p}\sim$0.097 shows the AFM order with $M_{\rm AFM}$(IPs)$\sim$0.1$\mu_{\rm B}$~\cite{Mukuda_review}. 

%fig4------------------------------------------------
\begin{figure}[tbp]
\centering
\includegraphics[width=6cm]{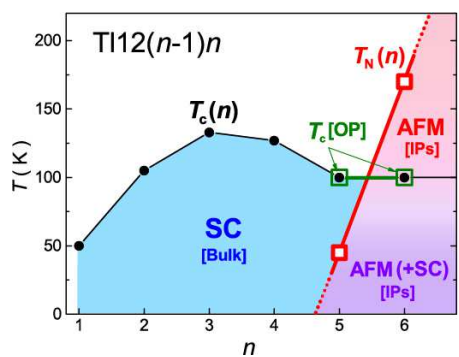}
\caption{(color online)  
Plots of $T_{\rm c}$($n$)\cite{Scott,Iyo_TcVsn} and $T_{\rm N}$($n$) as functions of $n$ for Tl12($n$-1)$n$ in optimally doped regime. 
%A legend of ${\bf hybrid AFM/SC}$ represents the coexisting state of AFM and SC at IPs to removes their misunderstanding that IPs are only responsible for the onset of AFM order, but not for the onset of SC order-parameter amplitude. 
In this series, the AFM order appears at the  IPs for $n\ge5$. 
Previous studies on $n$=5 revealed that the AFM order at the IPs  coexists uniformly below $T_{\rm N}$=45 K with the SC state below $T_{\rm c}'$[IPs]=87 K\cite{Mukuda2008,Mukuda_review}, enabling us to deduce that  the ground state of the IPs of $n$=6  is the uniform coexistence  of the AFM order with possible SC. 
}
\label{f4}
\end{figure}
%------------------------------------------------

Figure~\ref{f4} presents plots of $T_{\rm c}$($n$)~\cite{Scott,Iyo_TcVsn} and $T_{\rm N}$($n$)  for the homologous series of $n$-layered Tl12($n$-1)$n$  in the optimally doped regime. 
In this series, the AFM order appears at the IPs for $n\ge5$ but not for $n\le$4.
The marked increases in $T_{\rm N}$($n$) and $M_{\rm AFM}$($n$) from $n$=5 to 6 are considered to be associated with the increase in the interlayer magnetic coupling $J_{\rm out}$($n$). 
Note that when the total charge of the CRL is significantly reduced in the multilayered cuprates with apical fluorine, the AFM order appears for $n$=3 and 4~\cite{Shimizu_n3,Shimizu_n4}. 
$M_{\rm AFM}$($n$) at the AFM IPs in these compounds is much smaller than 0.5$\mu_{\rm B}$ in the Mott insulator CaCuO$_2$($n$=$\infty$)\cite{Tranquada}, since it is uniformly reduced by the mobile holes with $p$(IPs), which are responsible for the onset of SC order in the AFM {\it metallic} state. 
Here, we should emphasize that the AFM order at each IP in Tl1245($n$=5) coexists uniformly below $T_{\rm N}$=45 K with the SC state below $T_{\rm c}'$[IPs]=87 K\cite{Mukuda2008,Mukuda_review},  enabling us to deduce that the AFM order at the IPs of Tl1256($n$=6) also coexists with the possible SC order inherent to the IPs in the ground state.
%which removes their misunderstanding that IPs are only responsible for the onset of AFM order, but not for the onset of SC order-parameter amplitude in Tl1245($n$=5)\cite{Mukuda2008}. 
%In this context, both the AFM and SC orders are anticipated to coexist uniformly at IPs in Tl1256($n$=6) as well.

On the other hand, $T_{\rm c}\sim$100 K is maintained for $n$=5 and 6; nevertheless, 
%$T_{\rm N}$($n$) is by four times larger than $T_{\rm N}$($n$) and hence 
$T_{\rm N}$(6) ($T_{\rm N}$(5)) is significantly larger (smaller) than the bulk $T_{\rm c}$.
%, as seen in Fig. \ref{f4}.  
Here, it is interesting that $T_{\rm c}$$\sim$100 K for $n$=5 and 6 is also comparable with that for optimally doped single-layered Hg-1201,  indicating that the SC order parameter for $n$=5 and 6 is mostly dominated by the hole density at the optimally doped OP and the in-plane SC correlation. 
This is in contrast to the case of $n$=3, where a higher $T_{\rm c}$ ($\sim$133 K) is realized at three SC layers where the charge imbalance between the IP and OP is quite small\cite{Kotegawa2001}.
%at may be pair hopping between IP and OP due to the less difference of hole density.  
Note that the SC Cooper pairs in $n$=6 can tunnel between the OP through the AFM IPs by Josephson coupling as predicted in a theoretical study of multilayered cuprates\cite{Mori}.
In this context, we suggest that the possible interlayer Josephson coupling between the AFM IPs and the OP is insensitive to the magnitude of $T_{\rm N}$ and the AFM moments at the IPs or the AFM interaction among the IPs, demonstrating a novel interplay between the SC and AFM orders at $n\ge$5, where the charge imbalance between the IP and OP is significantly large. 

%Such an unique interplay of  the SC and AFM orders in the multilayered structure 
%,  that is,  the robust inplane SC order at OP with Tc =100K 
%on firm AFM order driven by  IPs with higher $T_{\rm N}$  than $T_{\rm c}$ that is  uniformly coexistence with the possible SC order at IPs
%The present result on $n=6$ reveals that the firm AFM order with higher $T_{\rm N}$  than $T_{\rm c}$ does not prevent the SC order in optimally-doped regime even when they are separated only in the atomic scale.  
%Robust inplane SC order on adjacent AFM order may be stemming from the ground state of uniformly coexistence of AFM order and SC order

%\section{Summary}===========================

In summary,  site-selective $^{63}$Cu-NMR/NQR and $^{205}$Tl-NMR studies on six-layered Tl1256 with $T_{\rm c}$=100 K have revealed that the AFM order with  moments of $0.10$ and $0.17\mu_{\rm B}$  takes place at the inner CuO$_2$ planes IP(1) and IP(2), respectively, below 170 K. 
This is the consequence of the underdoped hole densities at IP(1) and IP(2) with $p\sim$0.086 and $\sim$0.07, respectively. 
We highlight the fact that the increase in the out-of-plane magnetic interaction within the intra-unit-cell causes $T_{\rm N}(5)\sim$~45 K for Tl1245 with $n$(IP)=3 to increase to $T_{\rm N}(6)\sim$~170 K  for Tl1256 with $n$(IP)=4, whereas no trace of AFM order was observed for  $n\le 4$.  
The interesting finding in this work is that the marked increases in $T_{\rm N}$ and AFM moments for the IPs do not bring about any reduction in $T_{\rm c}$, since $T_{\rm c}\sim 100$~K is maintained for both the compounds with a nearly optimally doped OP. 
%It should be noted that the SC at $n$=6 occurs well below $T_{\rm N}$(6),  whereas the SC at $n$=5 does above $T_{\rm N}$(5).
%We suggest that a possible interlayer Josephson coupling between AFM IPs and OP is insensitive to the magnitude of $T_{\rm N}$ and AFM moments at IPs or the AFM interaction among IPs, demonstrating the intimate interplay between the SC and AFM orders in the doped Mott insulators.
We remark that even in the multilayered cuprates, the SC order for $n\ge5$ is mostly dominated by the long-range in-plane SC correlation, which is insensitive to the magnitude of $T_{\rm N}$ and the AFM moments at the IPs or the AFM interaction among the IPs, demonstrating the novel interplay between the SC and AFM orders at $n\ge5$. 
This is in contrast to the case of $n$=3, for which a higher $T_{\rm c}$ ($\sim$133 K) is realized at the three SC layers, where the charge imbalance between the IP and OP is quite small.

%\section*{Acknowledgements}

{\footnotesize 
This work was supported by JSPS KAKENHI Grant Nos. 26400356, 26610102, and 16H04013. }

%::::::::::::::::bibliography::::::::::::::::::::::::::::::::::::::::::::::::
%:::::::::::::::::::::::::::::::::::::::::::::::::::::::

\end{document}